# Weakly-bound clusters of atmospheric molecules: infrared spectra and structural calculations of $(CO_2)_n$ - $(CO)_m$ - $(N_2)_p$, $(n, m, p)$ = (2, 1, 0), (2, 0, 1), (1, 2, 0), (1, 0, 2), (1, 1, 1), (1, 3, 0), (1, 0, 3), (1, 2, 1), (1, 1, 2)

A.J. Barclay,[a] A.R.W. McKellar,[b] A. Pietropolli Charmet,[c] and N. Moazzen-Ahmadi[a]

[a]*Department of Physics and Astronomy, University of Calgary, 2500 University Drive North West, Calgary, Alberta T2N 1N4, Canada*

[b]*National Research Council of Canada, Ottawa, Ontario K1A 0R6, Canada*

[c]*Dipartimento di Scienze Molecolari e Nanosistemi, Università Ca' Foscari Venezia, Via Torino 155, I-30172, Mestre, Venezia, Italy*

**Abstract**

Structural calculations and high-resolution infrared spectra are reported for trimers and tetramers containing $CO_2$ together with CO and/or $N_2$. Among the 9 clusters studied here, only $(CO_2)_2$-CO was previously observed by high-resolution spectroscopy. The spectra, which occur in the region of the $\nu_3$ fundamental of $CO_2$ ($\approx 2350$ cm$^{-1}$), were recorded using a tunable optical parametric oscillator source to probe a pulsed supersonic slit jet expansion. The trimers $(CO_2)_2$-CO and $(CO_2)_2$-$N_2$ have structures in which the CO or $N_2$ is aligned along the symmetry axis of a staggered side-by-side $CO_2$ dimer unit. The observation of two fundamental bands for $(CO_2)_2$-CO and $(CO_2)_2$-$N_2$ shows that this $CO_2$ dimer unit is non-planar, unlike $(CO_2)_2$ itself. For the trimers $CO_2$-$(CO)_2$ and $CO_2$-$(N_2)_2$, the CO or $N_2$ monomers occupy equivalent positions in the 'equatorial plane' of the $CO_2$, pointing toward its C atom. To form the tetramers $CO_2$-$(CO)_3$ and $CO_2$-$(N_2)_3$, a third CO or $N_2$ monomer is then added off to the 'side' of the first two. In the mixed tetramers $CO_2$-$(CO)_2$-$N_2$ and $CO_2$-CO-$(N_2)_2$, this 'side' position is taken by $N_2$ and not CO. In addition to the fundamental bands, combination bands are also observed for $(CO_2)_2$-CO, $CO_2$-$(CO)_2$, and $CO_2$-$(N_2)_2$, yielding some information about their low-frequency intermolecular vibrations.

## 1. Introduction

The detailed role of weakly-bound van der Waals dimers and larger clusters in the earth's atmosphere remains an important, challenging, and somewhat uncertain subject. On the one hand, the concentration of dimers under atmospheric conditions is quite small, and those of larger clusters even smaller. On the other hand, as pointed out by Frohman et al.,[1] it is possible that, for example, even a small number of clustered $CO_2$ molecules could affect the absorption of radiation in the region of the $CO_2$ $\nu_2$ band. At any rate, the importance of infrared pressure broadening for atmospheric transmission is undoubted, and by studying weakly-bound clusters we obtain detailed information on intermolecular forces which are essential for understanding line broadening effects.

In the present paper, we report structural calculations and high-resolution infrared spectra for a number of trimers and tetramers containing $CO_2$ together with CO and/or $N_2$. Before considering these clusters, we first summarize previous experimental work on the spectra of the dimers, $CO_2$-CO and $CO_2$-$N_2$. The former dimer was originally observed by microwave and infrared spectroscopy [2-6] and found to have structure which is planar, T-shaped, and C-bonded, with $CO_2$ forming the 'top' of the T, and CO the 'stem'. Later, spectra of a second isomer, $CO_2$-OC, were discovered.[6,7] It has a similar T-shaped structure, but with the CO flipped by 180º, making it O-bonded. In the case of $CO_2$-$N_2$, infrared spectra[8-10] established an analogous T-shaped structure, and microwave spectra1 yielded precise rotational and hyperfine splitting parameters. The effective intermolecular center of mass separations are 3.91 Å for $CO_2$-CO, 3.58 Å for $CO_2$-OC, and 3.73 Å for $CO_2$-$N_2$. Calculations show that $CO_2$-CO is significantly more strongly bound than $CO_2$-OC, a fact which is important to keep in mind when thinking about the larger clusters studied in this paper. Not surprisingly, $CO_2$-$N_2$ has an intermediate binding energy. Rather than having two distinct isomers, $CO_2$-$N_2$ has the possibility of interchange of the two N atoms, but the resulting tunneling shifts and splittings are very small, being only barely resolved in the microwave spectrum.[1]

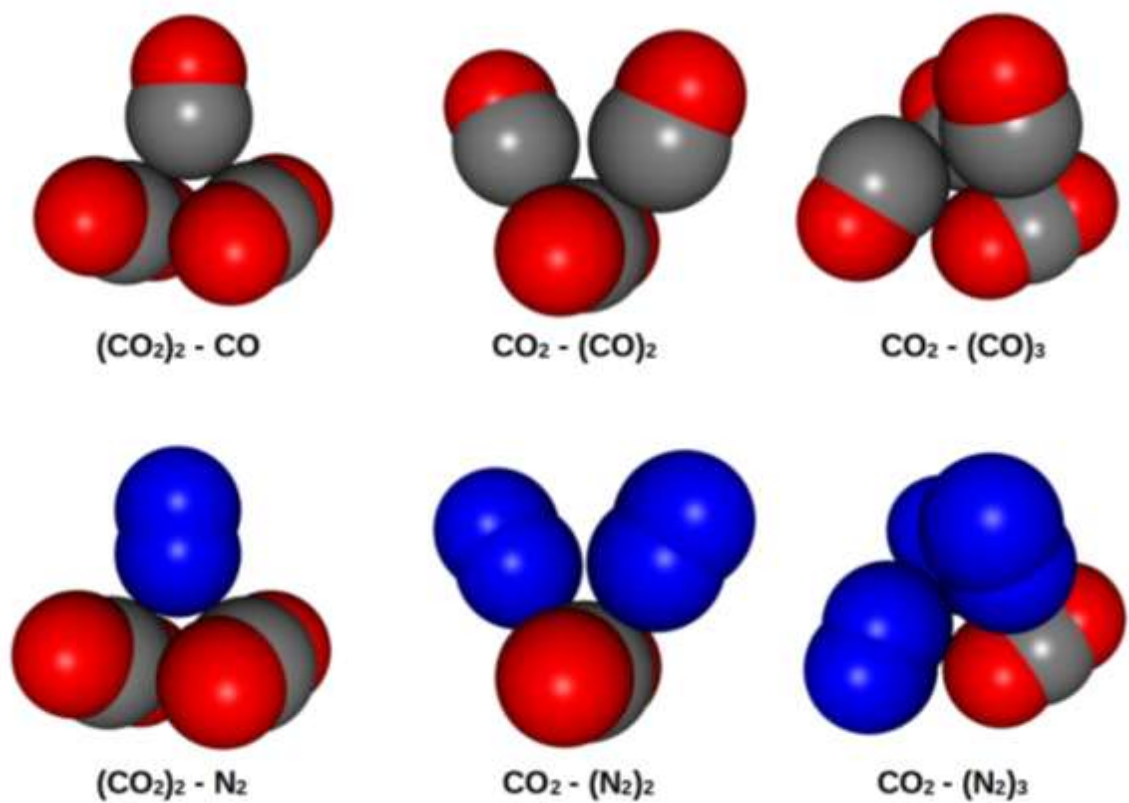

Fig. 1. Structures of some clusters studied here.

Some of the clusters studied here are illustrated in Fig. 1, and details of these structures are discussed later in this paper. We previously observed the spectrum of the $(CO_2)_2$-CO trimer in the region of the CO fundamental band ($\approx 2150$ cm$^{-1}$)[11] and found that its structure resembled that of a $CO_2$ dimer (planar slipped-parallel) with the CO monomer aligned along the dimer $C_2$ symmetry axis in a C-bonded configuration. The isolated $CO_2$ dimer is planar, but the planarity, or otherwise, of the dimer fragment in $(CO_2)_2$-CO could not be established at that time. In the present paper, we observe further $(CO_2)_2$-CO spectra, this time in the $CO_2$ $\nu_3$ region ($\approx 2350$ cm$^{-1}$), and show that the $CO_2$ dimer subunit does *not* remain planar within the trimer. As well, we observe spectra of the trimer $(CO_2)_2$-$N_2$ for the first time, and show that its structure is similar to that of $(CO_2)_2$-CO.

We observe fundamental and combination bands of $CO_2$-$(CO)_2$ and $CO_2$-$(N_2)_2$, neither of which has previously been studied by high resolution spectroscopy. Their structures have two equivalent CO or $N_2$ molecules located in the 'equatorial plane' of the $CO_2$, pointing approximately towards the C atom of the $CO_2$ and giving $C_{2v}$ symmetry. The position of each CO or $N_2$ relative to $CO_2$ is very similar to that in the T-shaped dimers mentioned above. These $C_{2v}$ trimer structures are analogous to that of $CO_2$-$Ar_2$.[12,13] The mixed trimer $CO_2$-CO-$N_2$ with a similar structure is also observed.

Finally, we observe the tetramers $CO_2$-$(CO)_3$ and $CO_2$-$(N_2)_3$. Their structures (Fig. 1) begin with the trimers from the previous paragraph and add another CO or $N_2$ off to the 'side' of the $CO_2$ 'equatorial plane'. There is a plane of symmetry ($C_s$ point group) which contains the $CO_2$ and the 'new' CO or $N_2$ molecule, and this plane bisects the angle between the first two CO or $N_2$. In addition, we observe band origins (though not rotational structure) for the mixed tetramers $CO_2$-$(CO)_2$-$N_2$ and $CO_2$-CO-$(N_2)_2$, and show that $N_2$ occupies the 'side' position in both cases.

There have been a number of theoretical studies of the $CO_2$ – CO interaction,[14,15] including two recent high level *ab initio* determinations of a four-dimensional potential surface.[16,17] In the case of $CO_2$-$N_2$, similar calculations have been reported at various levels of *ab initio* theory, most recently including a four-dimensional surface at the CCSD(T)-F12 triple zeta level.[18-20] Of course the structures of the present clusters also depend on the $CO_2$-$CO_2$, CO-CO, $N_2$-$N_2$, and CO-$N_2$ interaction potentials, which themselves have been extensively studied. Rather than relying on these various two-body potentials and their additivity, we report here new direct calculations for the clusters themselves in order to help confirm their structures.

**2. Structural calculations**

Since in this work we had to investigate many different clusters and deal with a large number of structures, we modified the computational approach followed in our previous investigations.[21,22] The present procedure is comprised of three distinct steps, each being carried out at a higher level of theory than the previous one.

In the first step, each single unit ($CO_2$, CO and $N_2$) of a given cluster was optimized as an isolated molecule at the lowest level of theory. Then, for the cluster under analysis, the sampling of its potential energy surface (PES) was performed by generating a large number (several hundred) of random initial structures, each then being optimized while treating all the molecules involved as rigid bodies (that is, only the orientation and position of the molecule in the resulting cluster could change). Redundant configurations were excluded, using as clustering criteria the mean squared deviations of their geometries and their energy difference.

In the second step, the remaining structures were optimized at an intermediate level of theory, but this time also allowing their intramolecular parameters to vary. As in the first step, optimized geometries which were identified as similar were excluded.

In the last step, each of the remaining unique structures obtained in the previous step was fully optimized at the highest level of theory. We also performed a frequency calculation to check that all these stationary points were real minima on the PES (i.e., no imaginary frequencies), and then computed their binding energies.

The first step employed the efficient (and very fast) semi-empirical GFN2-xtB[23] method using tight optimization criteria. For the second and third steps we used the B3LYP[24] and B2PLYP[25] functionals, respectively, since they have proven to compute rather accurate values of geometries and spectroscopic parameters.[26-28] The role of dispersion effects, crucial in case of calculations carried out with DFT methods,[29,30] was properly considered by means of the D3 corrections[31] and Becke-Johnson damping.[32] All the DFT calculations were carried out in conjunction with the m-aug-cc-pVTZ basis set[33] given its good performance demonstrated in our previous investigations on molecular clusters.

Accurate values of binding energies, extrapolated to the complete basis set limit (CBS) with the aug-cc-pV$n$Z basis sets ($n$ = D, T and Q),[34,35] were obtained for each of the optimized structures by means of the domain-based local pair natural orbital coupled cluster method (DLPNO-CCSD(T)),[36,37] which has been recently reported as a very efficient approach for the energetics of non-covalent structures.[38] All the DFT calculations have been carried out with the Gaussian16 software,[39] while the DLPNO-CCSD(T) computations have been performed with the Orca package.[40]

A summary of the results is given in Table I. For $(CO_2)_2$-CO, the most stable calculated isomer has rotational constants which agree well with the previous experimental results,[11] and the structure is similar to that deduced previously, combining the dimer $(CO_2)_2$ with a C-bonded CO molecule aligned along the dimer symmetry axis. However, the calculated $(CO_2)_2$ subunit is not planar, instead having each $CO_2$ with its axis tilted at 18.6° to the plane. Thus each $CO_2$ axis lies at 71.4° relative to the cluster symmetry axis (rather than 90° in the planar case), with the inner O atoms of each $CO_2$ pushed away from the CO molecule. The second most stable calculated $(CO_2)_2$-CO isomer is completely planar, effectively combining a distorted $(CO_2)_2$ with a distorted $CO_2$-CO. For $(CO_2)_2$-$N_2$, the most stable calculated isomer is analogous to that of $(CO_2)_2$-CO, with each $CO_2$ at an angle of 17.0° from planarity. The second calculated $(CO_2)_2$-$N_2$ isomer is a symmetric rotor having a central $N_2$, a linear C-N-N-C configuration, and crossed $CO_2$ molecules.

Table I. CBS extrapolated binding energies (BE, in kcal/mol) and rotational parameters ($A$, $B$, $C$, in cm$^{-1}$) for low lying isomers of clusters studied here. Values in parentheses include zero-point vibrational energy.

| Cluster, isomer # | BE | $A$ | $B$ | $C$ | Sym[a] | Description[b] |
|---|---|---|---|---|---|---|
| $(CO_2)_2$-CO #1 | -3.01(-2.31) | 0.0515 | 0.0485 | 0.0299 | $C_2$ | nonplanar $(CO_2)_2$, CO on symmetry axis |
| $(CO_2)_2$-CO #2 | -2.96(-2.28) | 0.0569 | 0.0387 | 0.0230 | $C_s$ | planar cluster |
| $(CO_2)_2$-$N_2$ #1 | -2.70(-2.08) | 0.0569 | 0.0480 | 0.0315 | $C_2$ | nonplanar $(CO_2)_2$, $N_2$ on symmetry axis |
| $(CO_2)_2$-$N_2$ #2 | -1.79(-1.29) | 0.1943 | 0.0133 | 0.0133 | $D_{2d}$ | symmetric top, linear C-N-N-C |
| $CO_2$-$(CO)_2$ #1 | -2.37(-1.74) | 0.0551 | 0.0505 | 0.0305 | $C_{2v}$ | 2 eq(C) |
| $CO_2$-$(CO)_2$ #2 | -2.11(-1.46) | 0.0638 | 0.0459 | 0.0267 | $C_s$ | 1 eq(C), 1 side, planar |
| $CO_2$-$(N_2)_2$ #1 | -1.99(-1.44) | 0.0620 | 0.0567 | 0.0349 | $C_{2v}$ | 2 eq |
| $CO_2$-$(N_2)_2$ #2 | -1.64(-1.15) | 0.0723 | 0.0469 | 0.0284 | $C_s$ | 1 eq, 1 side, planar |
| $CO_2$-CO-$N_2$ #1 | -2.16(-1.57) | 0.0584 | 0.0535 | 0.0326 | $C_s$ | eq(C) CO, eq $N_2$ |
| $CO_2$-CO-$N_2$ #2 | -1.93(-1.35) | 0.0665 | 0.0469 | 0.0275 | $C_s$ | eq(C) CO, side $N_2$, planar |
| $CO_2$-$(CO)_3$ #1 | -3.59 (-2.56) | 0.0323 | 0.0283 | 0.0271 | $C_s$ | 2 eq(C), 1 side (C) |
| $CO_2$-$(CO)_3$ #2 | -3.53 (-2.52) | 0.0552 | 0.0217 | 0.0169 | $C_s$ | 3 eq(C) (rotated) |
| $CO_2$-$(CO)_3$ #3 | -3.48 (-2.54) | 0.0348 | 0.0298 | 0.0283 | $C_s$ | 2 eq(C), 1 side (O) |
| $CO_2$-$(N_2)_3$ #1 | -3.03 (-2.21) | 0.0572 | 0.0244 | 0.0188 | $C_s$ | 3 eq (rotated) |
| $CO_2$-$(N_2)_3$ #2 | -2.99 (-2.15) | 0.0371 | 0.0301 | 0.0291 | $C_s$ | 2 eq, 1 side |
| $CO_2$-$(CO)_2$-$N_2$ #1 | -3.41(-2.47) | 0.0336 | 0.0286 | 0.0280 | $C_s$ | 2 eq(C) CO, side $N_2$ |
| $CO_2$-$(CO)_2$-$N_2$ #2 | -3.35(-2.41) | 0.0338 | 0.0292 | 0.0274 | $C_1$ | distorted-no symmetry plane, side CO |
| $CO_2$-$(CO)_2$-$N_2$ #3 | -3.34(-2.40) | 0.0548 | 0.0230 | 0.0177 | $C_s$ | 3 eq (rotated), CO center |
| $CO_2$-CO-$(N_2)_2$ #1 | -3.18(-2.29) | 0.0351 | 0.0298 | 0.0283 | $C_s$ | eq(C) CO, eq $N_2$, side $N_2$ |
| $CO_2$-CO-$(N_2)_2$ #2 | -3.17(-2.29) | 0.0575 | 0.0234 | 0.0182 | $C_s$ | 3 eq (rotated), $N_2$ center |
| $CO_2$-CO-$(N_2)_2$ #3 | -3.14(-2.23) | 0.0358 | 0.0295 | 0.0275 | $C_s$ | 2 eq $N_2$, side CO (C) |

[a] Point group symmetry.

[b] eq = 'equatorial' position; side = 'side' position; (C) = C-bonded; (O) = O-bonded.

The most stable calculated isomers of $CO_2$-$(CO)_2$ and $CO_2$-$(N_2)_2$ have similar structures, as shown in Fig. 1, with $C_{2v}$ symmetry and equivalent equatorial CO or $N_2$ molecules (C-bonded in the case of $CO_2$-$(CO)_2$). As shown below, the rotational constants agree quite well with experiment. The next most stable forms are planar, with one equatorial $N_2$ or C-bonded CO, and the other $N_2$ or CO located to the 'side'. In the case of $CO_2$-CO-$N_2$, the most stable form is analogous to $CO_2$-$(CO)_2$ and $CO_2$-$(N_2)_2$, with an equatorial $N_2$ and C-bonded CO. The second isomer is also analogous, and has an equatorial C-bonded CO with a 'side' mounted $N_2$.

Moving to the tetramers, we identified three low lying isomers for $CO_2$-$(CO)_3$ and two for $CO_2$-$(N_2)_3$, with results as shown in Table I. For $CO_2$-$(CO)_3$, isomer #1 has a geometry as shown in Fig. 1 and isomer #3 is similar, but with the 'side' CO unit flipped by roughly 180° so that it is O-bonded (but with the equatorial CO units still C-bonded). The structural similarity of isomers #1 and #3 is evident in their similar calculated rotational constants. Isomer #2 of $CO_2$-$(CO)_3$ has a different structure, with all three CO units located on the $CO_2$ 'equatorial plane' but with scrambled orientations, not simply pointing toward the C atom of the $CO_2$ as in $CO_2$-$(CO)_2$. Isomer #1 of $CO_2$-$(N_2)_3$ is similar to isomer #2 of $CO_2$-$(CO)_3$, while isomer #2 of $CO_2$-$(N_2)_3$ is similar to isomers #1 and #3 of $CO_2$-$(CO)_3$ with the third $N_2$ on the 'side' (of course there is no distinction between C- and O-bonded for $N_2$).

The lowest energy calculated isomer of $CO_2$-$(CO)_2$-$N_2$ has two equatorial C-bonded CO molecules and one 'side' located $N_2$. The lowest calculated isomer of $CO_2$-CO-$(N_2)_2$ has equatorial $N_2$ and C-bonded CO molecules, plus a 'side' mounted $N_2$, and the third isomer has a 'side' CO and two equatorial $N_2$. Meanwhile, the second isomer has all the CO and $N_2$ in the $CO_2$ equatorial plane, like isomer #2 of $CO_2$-$(CO)_3$. The preference in the calculations for $N_2$, rather than CO, to occupy the 'side' position in both $CO_2$-$(CO)_2$-$N_2$ and $CO_2$-CO-$(N_2)_2$ is confirmed by experiment as shown in Sec. 4.1 below.

## 3. Observed spectra and analysis

The spectra were recorded as described previously[41-43] using a rapid-scan optical parametric oscillator source to probe a pulsed supersonic slit jet expansion of dilute mixtures of $CO_2$ plus CO and/or $N_2$ in helium, with a backing pressure of about 11 atmospheres. Various mixtures were used, but for the spectra shown here, they contained about 0.05% $CO_2$ and 0.8 % $N_2$, about 0.1% $CO_2$ and 0.2% CO, or about 0.02% $CO_2$ and 0.8% CO. Wavenumber calibration was made using signals from a fixed etalon and a reference gas cell containing room temperature $CO_2$. Spectral simulation and fitting relied on the PGOPHER software,[44] and we generally used its Mergeblend option to fit blended lines to intensity weighted averages of their components.

### 3.1. Trimers $(CO_2)_2$-CO and $(CO_2)_2$-$N_2$

We previously analyzed a spectrum of $(CO_2)_2$-CO in the region of the CO fundamental band.[11] As a result, we started the current study with a reliable set of ground state rotational parameters which facilitated the search for further spectra in the $CO_2$ $\nu_3$ region. First to be recognized was a very weak combination band centered at 2375.78 cm$^{-1}$, as illustrated in Fig. 2. This band happens to lie just above a previously-studied[6] combination band of $CO_2$-CO. The new $(CO_2)_2$-CO combination band was relatively free from obscuration by other species, but even so assignment of such a weak and noisy spectrum would normally be very difficult. Once we realized that the band could be due to $(CO_2)_2$-CO, its successful analysis was aided by the known ground state parameters and by the visualization features of PGOPHER.

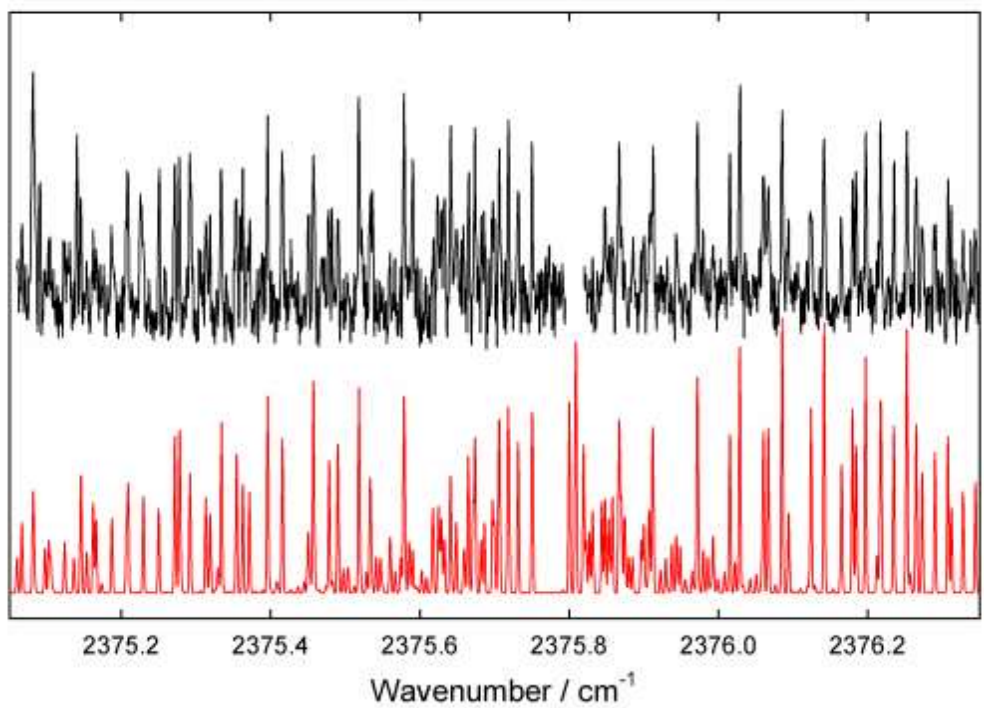

Fig. 2. Observed and simulated spectra of a combination band of $(CO_2)_2$-CO. Assignment of this very weak band was possible because ground state rotational parameters were already known. Stronger lines below 2375.2 cm$^{-1}$ are due to a nearby combination band of $CO_2$-CO.[6]

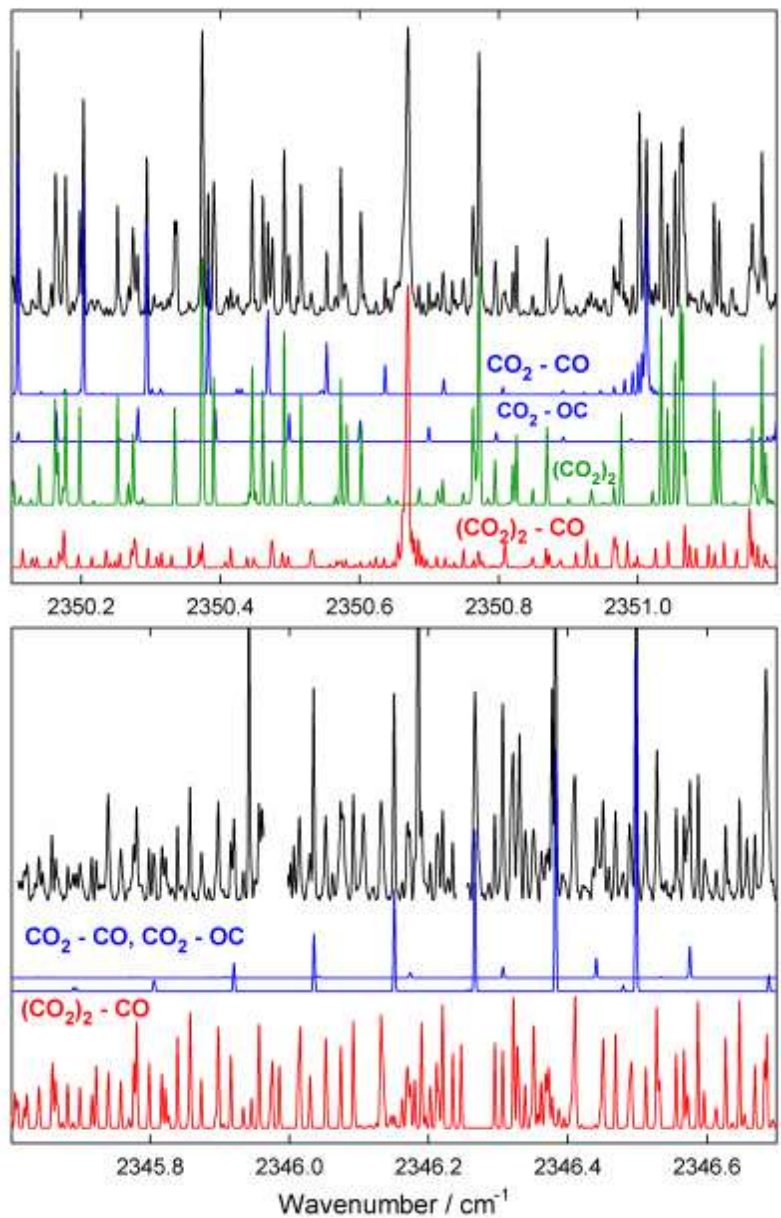

Fig. 3. Observed and simulated spectra of two fundamental bands of $(CO_2)_2$-CO. The vertical scale of the lower panel is magnified by a factor of 30 relative to the upper panel. Detection of the weak *b*-type fundamental in the lower panel shows that the $(CO_2)_2$ subunit within $(CO_2)_2$-CO does not remain planar as it is in the free $CO_2$ dimer.

Next, the assignment of a fairly strong *Q*-branch feature at 2350.67 cm$^{-1}$ (see Fig. 3) to $(CO_2)_2$-CO was confirmed by assigning the much weaker *P*- and *R*-branch transitions which accompany it. Here we had the problem of interference from many overlapping lines due to $(CO_2)_2$, $CO_2$-CO, $CO_2$-OC, and other unidentified species, but again had the advantage of known ground state parameters. This is a hybrid *a*- and *c*-type band arising from out-of-phase $\nu_3$ vibrations of the two $CO_2$ monomers, analogous to the well-known[45] fundamental band of $(CO_2)_2$ itself, whose origin is located nearby at 2350.77 cm$^{-1}$.

There are two $CO_2$ $\nu_3$ fundamental vibrations in $(CO_2)_2$, and in $(CO_2)_2$-CO, because each contains two $CO_2$ molecules. When the $CO_2$ molecules in $(CO_2)_2$ vibrate in-phase, their dipole transition moments cancel and transitions from the ground vibrational state have no intensity. This cancellation depends on the fact that $(CO_2)_2$ is planar, but the $(CO_2)_2$ subunit in $(CO_2)_2$-CO is not necessarily planar. In-phase vibrations in a nonplanar $(CO_2)_2$ subunit cancel in the (*a*, *c*) plane, but could give rise to *b*-type rotational transitions of $(CO_2)_2$-CO, and we discovered such a band centered at 2346.27 cm$^{-1}$, shown in the lower panel of Fig. 3. Again, assignment of this band would have been difficult without prior knowledge of the ground state parameters. The observation of the *b*-type fundamental shows that the $(CO_2)_2$ subunit is not planar, and more generally it further confirms assignment of these bands to $(CO_2)_2$-CO, rather than $CO_2$-$(CO)_2$ which has somewhat similar rotational constants (see below).

The $C_2$ symmetry axis of $(CO_2)_2$-CO means that only rotational levels with $(K_a, K_c)$ = (e,e) and (o,o) are allowed in the ground vibrational state.[11] We assigned 50 lines in the 2375.78 cm$^{-1}$ band, 27 in the 2350.77 cm$^{-1}$ band, and 70 in the 2346.27 cm$^{-1}$ band. Many lines were blends of more than one transition. The bands were analyzed simultaneously, including data from the previous 2150.59 cm$^{-1}$ band,[11] in order to obtain the best possible ground state parameters. Results are given in Table II. The root mean square (rms) errors were 0.00028, 0.00043, and 0.00028 cm$^{-1}$, respectively, for the three bands. The calculated rotational parameters for $(CO_2)_2$-CO in Table I agree very well with the experimental values in Table II. The fact that the calculated parameters are slightly ($\approx$1%) larger is expected since they are equilibrium values, while the experimental parameters include the effects of intermolecular zero-point motions.

The vertical scale in the bottom panel of Fig. 3 is magnified by a factor of about 30 relative to the upper panel, so the 2346.27 cm$^{-1}$ band is significantly weaker than the 2350.67 cm$^{-1}$ band. We estimated that the *b*-type dipole transition moment of the 2346.27 cm$^{-1}$ band was very roughly 0.3 times that of the total *a*- and *c*-type transition moments of the 2350.77 cm$^{-1}$ band, which themselves seemed to be roughly equal. This information is used below to help refine the experimental $(CO_2)_2$-CO structure. The 2375.78 cm$^{-1}$ combination band was weaker still, with an estimated relative transition moment of 0.1 compared to the 2350.77 cm$^{-1}$ band. These intensity estimates are very approximate, especially the latter, due to experimental variations in laser power and supersonic jet conditions.

We also assigned two fundamental bands of $(CO_2)_2$-$N_2$, aided by the analogy with $(CO_2)_2$-CO. Their relative strengths were similar to $(CO_2)_2$-CO, with the transition moment of the *b*-type band at 2346.39 cm$^{-1}$ being roughly 0.3 times that of the (*a*-, *c*)-type band at 2351.09 cm$^{-1}$. The *P*-branch region of the *b*-type band is illustrated in Fig. 4. The two bands were analyzed simultaneously with results as shown in Table II. A total of 64 lines of the *b*-type band and 22 lines of the (*a*-, *c*)-type band were fitted with an overall rms error of 0.00046 cm$^{-1}$. Experimental conditions for $(CO_2)_2$-$N_2$ were not as favorable as for $(CO_2)_2$-CO, so the $(CO_2)_2$-$N_2$ parameters are less well determined. So far, no combination bands have been observed for $(CO_2)_2$-$N_2$.

Table II. Molecular parameters for $(CO_2)_2$-CO and $(CO_2)_2$-$N_2$ (in cm$^{-1}$) [a]

|  | $(CO_2)_2$-CO | $(CO_2)_2$-$N_2$ |
|---|---|---|
| $A''$ | 0.0509267(53) | 0.056028(15) |
| $B''$ | 0.0482711(58) | 0.047728(17) |
| $C''$ | 0.0295110(23) | 0.0309468(58) |
| $\nu_0$, fund. 1 | 2346.2711(1) | 2346.3903(1) |
| $A'$ | 0.050845(14) | 0.055859(12) |
| $B'$ | 0.048302(11) | 0.047806(18) |
| $C'$ | 0.0295566(26) | 0.0309994(58) |
| $\nu_0$, fund. 2 | 2350.6716(1) | 2351.0851(2) |
| $A'$ | 0.050980(14) | 0.056143(18) |
| $B'$ | 0.048143(17) | 0.047541(23) |
| $C'$ | 0.0294498(48) | 0.030870(12) |
| $\nu_0$, comb. | 2375.7779(1) |  |
| $A'$ | 0.050833(10) |  |
| $B'$ | 0.047615(10) |  |
| $C'$ | 0.0292874(35) |  |

[a] Quantities in parentheses are 1$\sigma$ from the least-squares fit, in units of the last quoted digit.

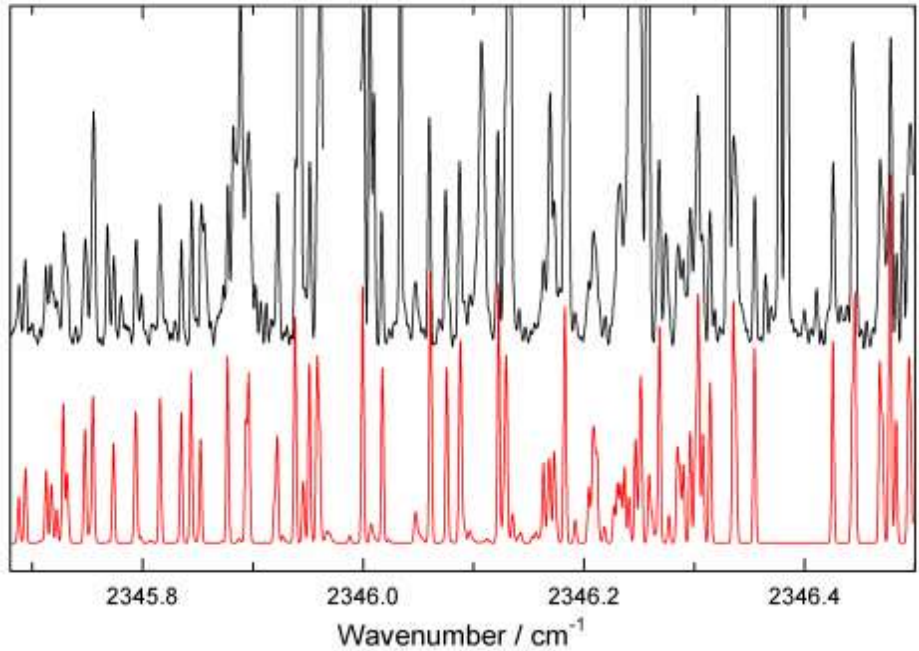

Fig. 4. Observed and simulated spectra showing the *P*-branch region of the *b*-type fundamental band of $(CO_2)_2$-$N_2$ (the band origin is at 2346.39 cm$^{-1}$). This is the clearest portion of our observed $(CO_2)_2$-$N_2$ spectra, but there are still many stronger overlapping lines, most of which are previously assigned (to $CO_2$-$N_2$, $CO_2$-He, etc.).

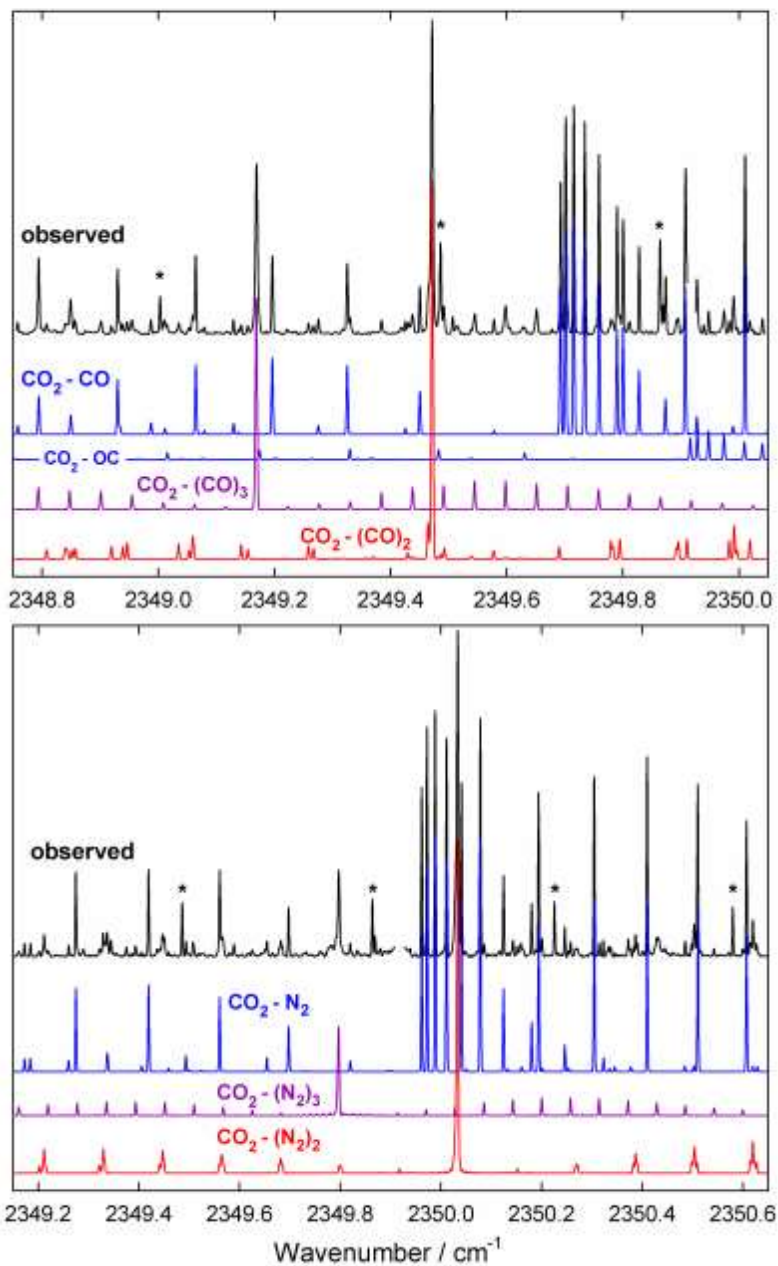

Fig. 5. Observed and simulated spectra of the fundamental bands of $CO_2$-$(CO)_2$ and $CO_2$-$(CO)_3$ (upper panel) and $CO_2$-$(N_2)_2$ and $CO_2$-$(N_2)_3$ (lower panel). Asterisks mark known transitions of $CO_2$-He.

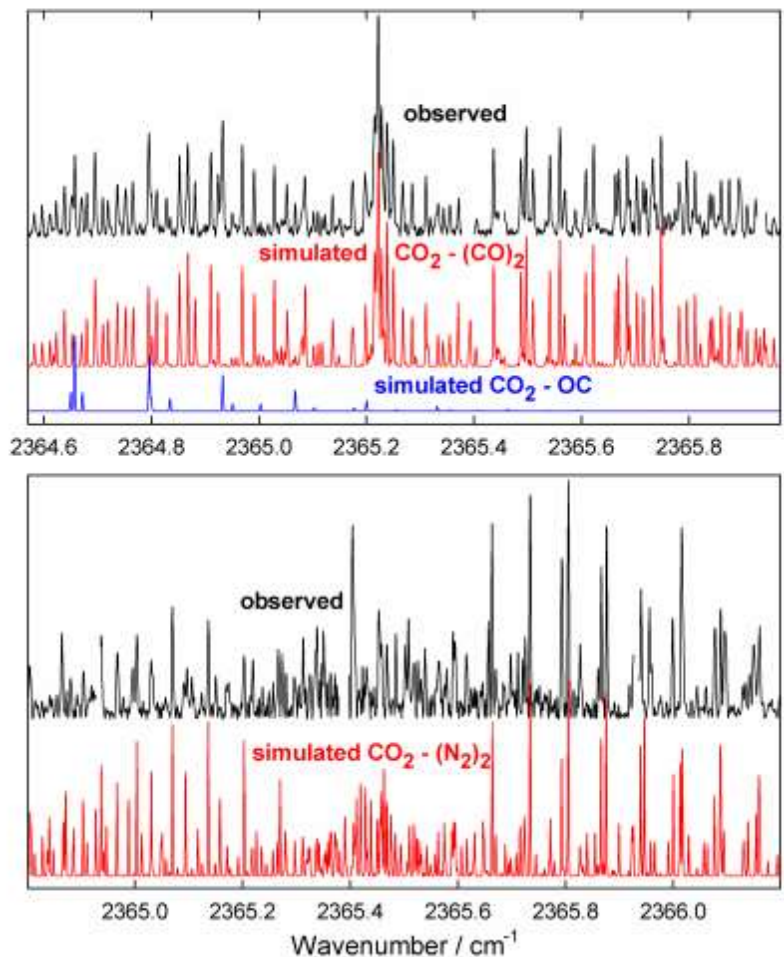

Fig. 6. Observed and simulated combination bands of $CO_2$-$(CO)_2$ and $CO_2$-$(N_2)_2$. The $CO_2$-$(CO)_2$ band lies just above a previously studied[6] combination band of $CO_2$-OC. The $CO_2$-$(N_2)_2$ band is highly perturbed, so only certain parts of the simulation match the observed spectrum well.

### 3.2. Trimers $CO_2$-$(CO)_2$, $CO_2$-$(N_2)_2$, and $CO_2$-CO-$N_2$

The first clue to the presence of the $CO_2$-$(CO)_2$ trimer was a fairly strong and otherwise unexplained $Q$-branch feature at 2349.47 cm$^{-1}$ in $CO_2$ + CO spectra (Fig. 5, top panel). A similar feature due to $CO_2$-$(N_2)_2$ appeared at 2350.04 cm$^{-1}$ in $CO_2$ + $N_2$ spectra (Fig. 5, bottom panel). With the help of the calculated structures from Sec. 2 it was possible to rotationally assign weaker $P$- and *R*-branch transitions accompanying these *Q*-branches. We also detected *a*-type combination bands of $CO_2$-$(CO)_2$ and $CO_2$-$(N_2)_2$ at 2365.23 and 2365.43 cm$^{-1}$, respectively, shown in Fig. 6. The $CO_2$-$(CO)_2$ combination band lies just above a known[6] combination band of $CO_2$-OC, and its analysis confirmed that of the 2349.47 cm$^{-1}$ fundamental band, including the observation that only levels with $(K_a, K_c)$ = (e,e) and (o,o) are populated in the ground vibrational state. This indicates that the two CO molecules within $CO_2$-$(CO)_2$ are indeed equivalent. We assigned 53 lines in the fundamental band and 83 lines in the combination band. These were fit simultaneously with rms errors of 0.00023 and 0.00040 cm$^{-1}$, respectively. Resulting parameters are given in Table III for both bands. There was no evidence of singly or doubly O-bonded forms of $CO_2$-$(CO)_2$, but these might still exist as stable isomers.

Analysis of the $CO_2$-$(N_2)_2$ fundamental band at 2350.04 cm$^{-1}$ (Fig. 5, bottom panel) was similar to that of $CO_2$-$(CO)_2$ except for spin statistics. If the two $N_2$ units in $CO_2$-$(N_2)_2$ are equivalent, then the spin weights are 5:4 for levels with $(K_a, K_c)$ = (e,e), (o,o) : (e,o), (o,e) (the symmetry axis is *b*), as compared to 1:0 for $CO_2$-$(CO)_2$. However, this slight difference was not noticeable in our spectra. Double the number of transitions, compared to $CO_2$-$(CO)_2$, made the $CO_2$-$(N_2)_2$ analysis more difficult and less precise. The *a*-type $CO_2$-$(N_2)_2$ combination band (Fig. 6, bottom panel) turned out to be highly perturbed. Transitions with $K_a = 0$ and 1 could be assigned and confirmed by ground state combinations differences, but assignments for higher $K_a$ were increasingly uncertain. We were not able to explain the perturbations with a simple model involving a single perturbing state.

Fitting the *c*-type $CO_2$-$(N_2)_2$ fundamental band by itself resulted in a relatively large uncertainty for the *C* rotational constant. In order to utilize the information contained in the perturbed combination band (particularly for *C*), we decided to determine the ground state parameters using ground state combination differences derived from both bands. Then the infrared bands were fit using these fixed ground state parameters. The ground state analysis involved 24 unique combination differences (from 47 lines) which were fit with an rms error of 0.00041 cm$^{-1}$. For the fundamental band, 57 lines were then fit with an rms error of 0.00044 cm$^{-1}$. For the combination band, the fit included only a limited number of 28 transitions and the rms error was 0.00080 cm$^{-1}$. But the resulting combination band parameters and their uncertainties may not be very significant since only those transitions which fit well were included. The $CO_2$-$(N_2)_2$ parameters are given in Table III. Note that the rotational constants are larger than those of $CO_2$-$(CO)_2$ by 6% to 20%, just as the rotational constants of $CO_2$-$N_2$ are larger than those of $CO_2$-$CO$.

Using a mixture containing $CO_2$, $CO$, and $N_2$ in helium, a new $Q$-branch feature appeared at 2349.75 cm$^{-1}$, almost exactly midway between those of $CO_2$-$(CO)_2$ and $CO_2$-$(N_2)_2$. We naturally assigned this feature to the mixed trimer $CO_2$-$CO$-$N_2$. By carefully comparing the new spectrum with those involving just $CO_2$+$CO$ or $CO_2$+$N_2$, it was possible to assign a number of $P$- and $R$-branch transitions to $CO_2$-$CO$-$N_2$, helped by the fact that the rotational parameters of the new trimer were, as expected, close to midway between those of $CO_2$-$(CO)_2$ and $CO_2$-$(N_2)_2$. We assigned 28 lines and fitted them with an rms error of 0.00039 cm$^{-1}$ to obtain the parameters given in Table III. In this fit, the $C$ parameter was not very well determined, even when constrained to be equal in the ground and excited states.

The calculated rotational parameters for $CO_2$-$(CO)_2$, $CO_2$-$(N_2)_2$, and $CO_2$-$CO$-$N_2$ in Table I agree fairly well with the experimental values in Table III. However, the agreement is not as good as for $(CO_2)_2$-$CO$, perhaps because the effects of large amplitude motions are more important for the $CO_2$-$(CO)_2$ family.

Table III. Molecular parameters for $CO_2$-$(CO)_2$, $CO_2$-$(N_2)_2$, and $CO_2$-CO-$N_2$ (in cm$^{-1}$). [a]

|  | $CO_2$-$(CO)_2$ | $CO_2$-$(N_2)_2$ | $CO_2$-CO-$N_2$ |
|---|---|---|---|
| $A''$ | 0.056541(21) | 0.060054(23) | 0.058395(29) |
| $B''$ | 0.049312(12) | 0.057499(21) | 0.052808(34) |
| $C''$ | 0.030127(10) | 0.034194(14) | 0.03124(56) |
| $10^6 \times D_K$ | 1.83(59) | -1.50(73) | [0.0] |
| $10^6 \times D_{JK}$ | 1.29(46) | 3.87(74) | [0.0] |
| $10^7 \times D_J$ | 1.44(63) | -2.4(11) | [0.0] |
| $\nu_0$, fund. [b] | 2349.4721(1) | 2350.0351(1) | 2349.7536(1) |
| $A'$ | 0.056447(25) | 0.0600096(24) | 0.058261(25) |
| $B'$ | 0.049294(13) | 0.0573571(25) | 0.052761(28) |
| $C'$ | 0.030134(15) | 0.0341670(70) | [0.03124] [c] |
| $\nu_0$, comb. [d] | 2365.2301(1) | 2365.4296(3) |  |
| $A'$ | 0.056100(24) | 0.06013(11) |  |
| $B'$ | 0.050318(15) | 0.06009(9) |  |
| $C'$ | 0.0302874(71) | 0.034277(6) |  |

[a] Quantities in parentheses are 1σ from the least-squares fit, in units of the last quoted digit. The $CO_2$-$(N_2)_2$ ground state fit was made to combination differences (see text). The $CO_2$-$(N_2)_2$ excited state fits were made with fixed ground state parameters. The $CO_2$-$(N_2)_2$ combination band is highly perturbed, so its parameters have limited significance.

[b] The fundamental excited state centrifugal distortion parameters were constrained to equal the ground state values.

[c] Excited state parameter constrained to equal ground state value.

[d] The combination band excited state centrifugal distortion parameters were fixed to zero, except $D_{JK} = 2.49(33) \times 10^{-6}$ cm$^{-1}$ for $CO_2$-$(CO)_2$.

Table IV. Molecular parameters for $CO_2$-$(CO)_3$ and $CO_2$-$(N_2)_3$ (in cm$^{-1}$) [a]

|  | $CO_2$-$(CO)_3$ ground state | $CO_2$-$(CO)_3$ fundamental | $CO_2$-$(N_2)_3$ ground state | $CO_2$-$(N_2)_3$ fundamental |
|---|---|---|---|---|
| $\nu_0$ |  | 2349.1700(2) |  | 2349.7988(4) |
| $A$ | [0.03131] [b] | 0.03129(2) | [0.03613] [b] | 0.03606(4) |
| $(B + C)/2$ | 0.0268440(71) | 0.0268255(76) | 0.028823(11) | 0.028793(10) |
| $(B - C)$ | [0.0005] | [0.0005] | [0.0005] | [0.0005] |

[a] Quantities in parentheses correspond to $1\sigma$ from the least-squares fit, in units of the last quoted digit. Quantities in square brackets were fixed in the fits.

[b] These are fixed equal to the calculated $A$-values (Table I), scaled by the ratios of the experimental and calculated $(B + C)/2$ values.

### 3.3. Tetramers $CO_2$-$(CO)_3$, $CO_2$-$(N_2)_3$, $CO_2$-$(CO)_2$-$N_2$, and $CO_2$-CO-$(N_2)_2$

Additional $Q$-branch features were evident in Fig. 5 in the $CO_2$ + CO (2349.17 cm$^{-1}$) and $CO_2$ + $N_2$ (2349.80 cm$^{-1}$) spectra. Each $Q$-branch was surrounded by simple, regularly spaced series of $P$- and $R$-transitions, giving the appearance of parallel bands of a symmetric or near-symmetric rotor. We believe that these new bands are due to the tetramers $CO_2$-$(CO)_3$ and $CO_2$-$(N_2)_3$. This assignment was supported by examining the relative intensities of various bands in spectra with different concentration ratios of $CO_2$ and CO.

We assigned 13 blended lines of $CO_2$-$(CO)_3$ (including the $Q$-branch) which were fitted with an rms error of 0.00041 cm$^{-1}$. Similarly, for $CO_2$-$(N_2)_3$ 15 blended lines were fitted with an error of 0.00072 cm$^{-1}$. These spectra have only limited information content. As shown in Table IV, four parameters were varied for each tetramer: the band origin, $(B + C)/2$ for the ground and excited state, and the change in $A$-value between ground and excited states. The band origins and $(B + C)/2$ values were well determined, but the observed spectra were not sensitive to the value of *A*, and also not sensitive to (*B* - *C*) as long as it was less than about 0.0008 cm$^{-1}$.

The structure of the tetramers was not obvious from the spectra, and we turned for guidance to the calculations described in Sec. 2, which gave structures similar to those in Fig. 1 with the third CO or $N_2$ molecule located to the 'side' of the first two. These are not symmetric rotors, but their basic geometry can easily give 'accidental' near-symmetric rotors with the observed parallel band structure, as discussed further in Sec. 4.2.3.

In the combined $CO_2$ + CO + $N_2$ spectrum mentioned above, two new peaks appeared in addition to the 2349.75 cm$^{-1}$ peak assigned above to $CO_2$-CO-$N_2$. We believe that a peak at 2349.518 cm$^{-1}$ is due to $CO_2$-CO-$(N_2)_2$, and that one at 2349.253 cm$^{-1}$ is due to $CO_2$-$(CO)_2$-$N_2$, and that these tetramers have structures like $CO_2$-$(CO)_3$ and $CO_2$-$(N_2)_3$, with two equatorial and one 'side' CO or $N_2$. Detailed rotational analysis of these mixed tetramers was not possible. As discussed below in Sec. 4.1, the observed vibrational shifts indicate that the 'side' position is occupied by $N_2$ in both cases. The other isomers, with CO in the side position, are not evident in the spectrum. This agrees with our calculations (Table I) which indicate that $N_2$ is energetically favored in the side position.

## 4. Discussion

### 4.1. Vibrational shifts

The 2346.271 cm$^{-1}$ fundamental of $(CO_2)_2$-CO corresponds to the infrared forbidden $A_g$ fundamental of $(CO_2)_2$. This $(CO_2)_2$ frequency is not known experimentally, but has been estimated to be 2346.76 cm$^{-1}$ by modeling $(CO_2)_2$ isotope effects.[46] If this estimate is correct then the vibrational shift in $(CO_2)_2$-CO relative to $(CO_2)_2$ is about -0.49 cm$^{-1}$. The 2350.672 cm$^{-1}$ fundamental of $(CO_2)_2$-CO represents a shift of -0.099 cm$^{-1}$ relative to the allowed $B_u$ fundamental of $(CO_2)_2$. For $(CO_2)_2$-$N_2$, the corresponding shifts are -0.37 cm$^{-1}$ for the $A_g$ mode and +0.413 cm$^{-1}$ for the $B_u$ mode. The fact that both are more positive (blue shifted) compared to those of $(CO_2)_2$-CO is consistent with the shifts observed for the dimers, where $CO_2$-$N_2$ (+0.484 cm$^{-1}$) is more blue-shifted than $CO_2$-CO (+0.211 cm$^{-1}$).

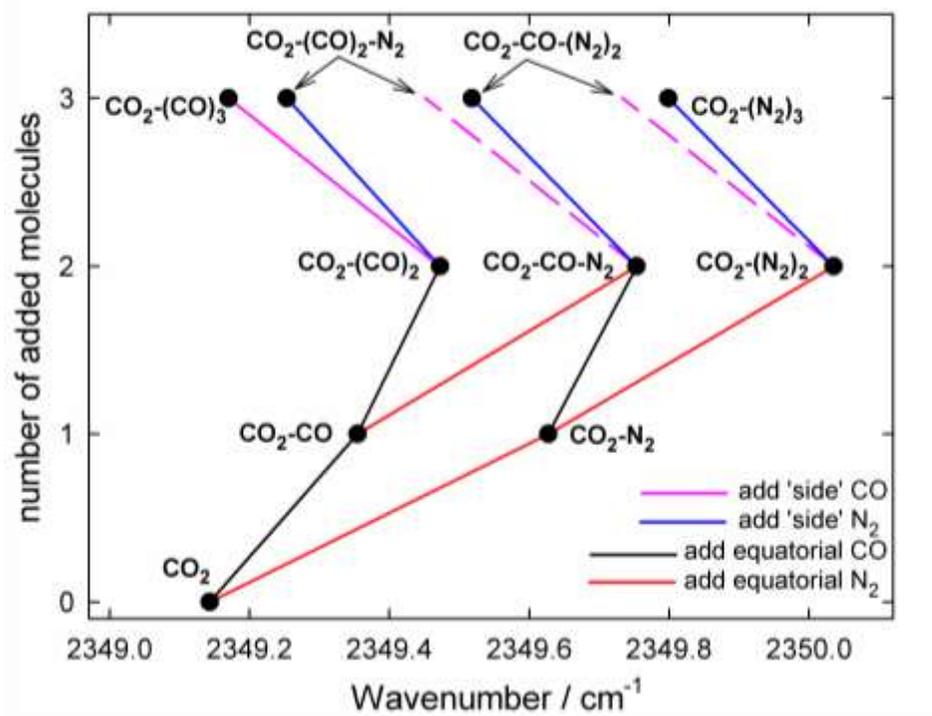

Fig. 7. Solid circles indicate observed band origins for $CO_2$-$(CO)_m$-$(N_2)_p$ clusters. Color coded lines show the vibrational shifts induced by adding CO or $N_2$ in equatorial or 'side' positions. The results strongly suggest that the observed mixed tetramers $CO_2$-$(CO)_2$-$N_2$ and $CO_2$-CO-$(N_2)_2$ have $N_2$ in the side position, not CO.

Vibrational origins of all the $CO_2$-$(CO)_m$-$(N_2)_p$ trimers and tetramers studied here are shown graphically in Fig. 7, together with the origins of the previously studied dimers and of $CO_2$ itself. The shifts for (m + p) = 1 and 2 progress in a reasonably linear fashion, which is understandable since the first and second added monomers occupy equivalent positions relative to $CO_2$, though the additional shift induced by the second CO or $N_2$ is a bit smaller than that induced by the first. The third CO or $N_2$ molecule has a rather different effect, inducing a red shift, and this difference is understandable because the position is different. The similarity of the shift patterns for CO and $N_2$ is evidence for the similarity of their cluster structures.

The points in Fig. 7 at 2349.253 and 2349.518 cm$^{-1}$ labeled as $CO_2$-$(CO)_2$-$N_2$ and $CO_2$-CO-$(N_2)_2$ represent the unresolved $Q$-branches mentioned above (Sec. 3.3). There are two ways to form the tetramer $CO_2$-$(CO)_2$-$N_2$ from a trimer: either add CO in the 'side' position to $CO_2$-CO-$N_2$, or else add $N_2$ in the side position to $CO_2$-$(CO)_2$. In the former case (side CO), we observe nothing at the expected position of about 2349.45 cm$^{-1}$. But in the latter case (side $N_2$) we do observe the expected $Q$-branch at 2349.253 cm$^{-1}$. A similar argument applies to $CO_2$-CO-$(N_2)_2$, for which we assign the 2349.518 cm$^{-1}$ $Q$-branch to the isomer with $N_2$ in the side position, but observe nothing around 2349.73 cm$^{-1}$, the expected origin for CO were in the side position. This preference for $N_2$ to occupy the 'side' position agrees with our calculations (Sec. 2 and Table I).

### 4.2. Experimental structures

#### 4.2.1. $(CO_2)_2$-CO and $(CO_2)_2$-$N_2$

We previously had ground state rotational parameters[11] for $(CO_2)_2$-CO and $(C^{18}O_2)_2$-CO, and now have one new piece of experimental information, namely detection of the *b*-type fundamental and its approximate relative intensity. The intensity measurement given above yields an estimate of $\sin^{-1}(0.3) = 17°$ for $\phi$, the departure of each $CO_2$ unit from planarity. This is close to our calculated angle of $\phi = 18.6°$ from Sec. 2. Assuming the experimental out-of-plane angle, the overall fit from Ref. 11 is slightly improved, and gives the following structural parameters: $R_1$ (center of mass (c.m.) separation of CO and $(CO_2)_2$ subunits) = 3.512(2) Å; $R_2$ (c.m. separation of $CO_2$ subunits) = 3.492(2) Å; and $\theta$ (angle between line connecting the C atoms of the $CO_2$ units and an OCO axis) = 60.7(2)°. For comparison, our theoretical equilibrium structure from Sec. 2 has $R_1$ = 3.460 Å, $R_2$ =

3.499 Å, and $\theta = 63.2°$, and the experimental structure of $(CO_2)_2$ itself has $\phi = 0°$, $R_2 = 3.60$ Å, and $\theta = 57.9°$. (The experimental structures assume that the CO and $CO_2$ monomers remain unchanged in the clusters.)

There are no isotopic data for $(CO_2)_2$-$N_2$, but using the ground state rotational parameters from Table II, and assuming the same value of $\phi = 17°$, we obtain $R_1 = 3.306$ Å, $R_2 = 3.519$ Å, and $\theta = 60.6°$, while the theoretical equilibrium structure from Sec. 2 has $R_1 = 3.251$ Å, $R_2 = 3.513$ Å, and $\theta = 62.6°$. We note that the experimental structural changes between $(CO_2)_2$-CO and $(CO_2)_2$-$N_2$ are very well predicted by theory.

### 4.2.2. $CO_2$-$(CO)_2$ and $CO_2$-$(N_2)_2$

The missing levels (due to nuclear spin statistics) observed for $CO_2$-$(CO)_2$ show that the two CO molecules are equivalent, and the same is probably true for $CO_2$-$(N_2)_2$, though we cannot be absolutely sure since the slight expected 5:4 intensity alternation is not detectable. It is still possible that the equilibrium structures could be slightly asymmetric, since we know that the $(CO)_2$ and $(N_2)_2$ dimers[47,48] prefer to have staggered rather than exactly side-by side monomers. But in that case the symmetric saddle-point in the potential must be low enough that the zero-point positions of the CO are effectively equivalent. In any case, symmetric structures are supported by our theoretical calculations in Sec. 2.

Given the symmetric $C_{2v}$ geometry and assuming the monomer structures remain unchanged, three parameters describe the structure of $CO_2$-$(CO)_2$ or $CO_2$-$(N_2)_2$. These include $R_1$, the c.m. distance from CO to $CO_2$, and $\theta$, the angle from one CO c.m. to the $CO_2$ c.m. to the other CO c.m. (together, these determine the CO to CO distance, $R_2$). In addition, since the CO molecules need not point directly at the $CO_2$, there is also a parameter $\phi$ describing the deviation from 180° of the angle subtended by O-C-C (or N-N-C), where the first two atoms belong to CO (or $N_2$) and the third to $CO_2$. One might expect that repulsion between the C atoms of the CO (or the inner N atoms of the $N_2$) would result in $\phi$ being less than 180°, making the CO or $N_2$ molecules more nearly parallel to each other. This is indeed the case for our calculated structures, which have $\phi = 173.8°$ for $CO_2$-$(CO)_2$, $\phi = 174.1°$ for $CO_2$-$(N_2)_2$, and similar angles of 173.8° and 173.7° for $CO_2$-CO-$N_2$.

When we try to determine a structure from the observed rotational constants, it turns out that the parameters ($R_1$, $\theta$, $\phi$) are highly correlated. Fixing $\phi$ at its calculated value of 173.8° gives an "experimental" $CO_2$-$(CO)_2$ structure with $R_1 = 3.90$ Å and $\theta = 72.4°$. This in turn gives $R_2 = 4.60$ Å, a C-C distance of 3.95 Å (CO to CO), an O-O distance of 5.08 Å (CO to CO), and a C-C distance of 3.26 Å (CO to $CO_2$). For comparison, our calculated equilibrium structure (Sec. 2) has $R_1 = 3.90$ Å and $\theta = 70.9°$. Using the calculated $\phi$ for $CO_2$-$(N_2)_2$ we obtain "experimental" values of $R_1 = 3.69$ Å and $\theta = 69.5°$. This results in $R_2 = 4.21$ Å, an inner N-N distance of 3.68 Å, an outer N-N distance of 4.74 Å, and an inner N-C distance of 3.14 Å. Our calculated equilibrium structure (Sec. 2) has $R_1 = 3.71$ Å and $\theta = 65.4°$. We did not try to determine an experimental structure for $CO_2$-CO-$N_2$ since the experimental rotational constants are not well determined and there are many free structural parameters.

The experimental $CO_2$ to CO or $N_2$ c.m. separations of 3.90 or 3.69 Å determined here for $CO_2$-$(CO)_2$ and $CO_2$-$(N_2)_2$ are similar to but slightly smaller than the values of 3.91 and 3.73 Å previously determined for $CO_2$-CO and $CO_2$-$N_2$.[2-7] We attribute this shrinkage to the general effects of anharmonicity, as zero-point motions tend to become smaller due to extra mass and bonding in the trimers compared to the dimers. But these are still weakly-bound systems with large amplitude motions, so we cannot expect to fully describe their structure and dynamics in terms of a simple fixed geometry. The fact that the $CO_2$ to CO separation for $CO_2$-$(CO)_2$ is similar to $CO_2$-CO, and not $CO_2$-OC, further supports assignment of the observed trimer as C-bonded and not O-bonded. It is still possible that singly and/or doubly O-bonded $CO_2$-$(CO)_2$ also exist as stable isomers, but we have not seen any experimental evidence for them.

### 4.2.3. $CO_2$-$(CO)_3$ and $CO_2$-$(N_2)_3$

It is not possible to establish simple fixed "experimental" structures for these tetramers, since each has only one well determined experimental rotational parameter, $(B + C)/2$, while many geometrical parameters are required to specify their structures. If we allow for the fact that the theoretical equilibrium structures are likely to overestimate experimental rotational constants (due to anharmonic effects), then calculated isomer #1 (or #3) of $CO_2$-$(CO)_3$ and isomer #2 of $CO_2$-$(N_2)_3$ from Table I provide quite good matches to the experimental $(B + C)/2$ values in Table IV, while the theoretical values of $(B - C)$ are somewhat too large. It is interesting to note that the $b$- and $c$-inertial axes interchange between isomers #1 and #3 of $CO_2$-$(CO)_3$, meaning that a 'blend' of these isomers could have a very small $(B - C)$ value. Such a 'blend' is possible if the 'side' CO unit is somewhat free to rotate. More generally, relatively small changes in the geometry of $CO_2$-$(CO)_3$ isomer #1 and $CO_2$-$(N_2)_3$ isomer #2 can have large effects on the $(B - C)$ value. In particular, the angle between one equivalent CO or $N_2$, the $CO_2$, and the other equivalent CO or $N_2$, called $\theta$ for the trimers above, has a large effect on $(B - C)$. Increasing this angle by only about 2° is sufficient to reduce $(B - C)$ to zero in $CO_2$-$(CO)_3$ and $CO_2$-$(N_2)_3$, and such an increase is similar to the difference between calculated and "experimental" $\theta$ values noted above for $CO_2$-$(CO)_2$ and $CO_2$-$(N_2)_2$. To summarize, we conclude that the observed spectra of $CO_2$-$(CO)_3$ and $CO_2$-$(N_2)_3$ are due to clusters with structures similar to those shown in Fig. 1 (isomers #1 and #2, respectively, from Table I).

There remains the problem that the observed form of $CO_2$-$(N_2)_3$ does not correspond to the most stable calculated one, but rather to the second most stable, even if the computed energy difference between them can be considered really very small. In contrast, the observed forms do agree with most stable isomer for all the other clusters in Table I. Could the tetramers with co-planar equatorial CO or $N_2$ molecules (isomer #2 of $CO_2$-$(CO)_3$ and isomer #1 of $CO_2$-$(N_2)_3$) also be present experimentally, but not observed yet? Or are they (relatively) a bit less strongly bound than indicated by the present calculations?

### 4.3. Combination bands

For $(CO_2)_2$-CO, we observe a combination band at 2375.78 cm$^{-1}$ with *b*-type selection rules (Fig. 2), which represents an intermolecular frequency of 29.507 cm$^{-1}$ if associated with the *b*-type fundamental 1, or 25.106 cm$^{-1}$ if associated with the (*a*-, *c*)-type fundamental 2. Within the $C_2$ point group, the symmetries of fundamental 1, fundamental 2, and the combination mode are A, B, and A, respectively. Thus the intermolecular mode itself must have A symmetry if associated with fundamental 1, or B if associated with fundamental 2.

$(CO_2)_2$-CO has nine intermolecular modes. Four of these are analogous to the intermolecular modes of $(CO_2)_2$ itself,[49,50] namely: $CO_2$ in-plane geared bend (B symmetry), $CO_2$ in-plane anti-geared bend (A), $CO_2$ torsion (A), $CO_2$-$CO_2$ van der Waals stretch (A). In $(CO_2)_2$ these have predicted[49] values of 20.6, 92.2, 24.4, and 46.1 cm$^{-1}$, respectively. Of course the $CO_2$ units in $(CO_2)_2$-CO are not planar, but their intermolecular modes should be similar and the given symmetries apply to $(CO_2)_2$-CO. Addition of the CO unit introduces five additional intermolecular modes: two geared bends (B), two antigeared bends (B), and a $CO_2$-CO van der Waals stretch (A). (The two types of each bend can be thought of as being parallel either to the *a*- or else *c*-axis of $(CO_2)_2$-CO.) So there are many possibilities for assigning the observed combination band! The more likely ones include: fundamental 1 plus $CO_2$ torsion, fundamental 2 plus $CO_2$ geared bend, or fundamental 2 plus one of the $CO_2$-CO geared bends. As a further possibility, we note that $(CO_2)_2$ has an observed combination band which likely corresponds to the twice the $CO_2$ geared bend, giving an intermolecular bending overtone frequency of 31.5 cm$^{-1}$.[50] The analogous overtone mode in $(CO_2)_2$-CO would have symmetry A, and could thus combine with fundamental 1 to give the observed (29.5 cm$^{-1}$) combination band.

For $CO_2$-$(CO)_2$, we observe the *a*-type combination band at 2365.23 cm$^{-1}$ (Fig. 6), which represents an intermolecular frequency of 15.758 cm$^{-1}$ with respect to the fundamental. $CO_2$-$(CO)_2$ has $C_{2v}$ symmetry, and we choose the axis system so that the *c*-type fundamental has $B_1$ symmetry and the combination band is $B_2$. This means that the relevant intermolecular mode must be $A_2$, since $B_1 \otimes A_2 = B_2$. Again, there are nine fundamental intermolecular modes, and here we focus on the two $A_2$ modes, which maintain the $C_2$ rotational symmetry of $C_{2v}$ but destroy its two symmetry planes. One $A_2$ mode can be described as a torsion in which the $CO_2$ twists around the $C_2$ symmetry axis with respect to the CO molecules, which remain coplanar. The other $A_2$ mode can be described as a torsion of the $(CO)_2$ subunit, with the $CO_2$ and CO centers of mass remaining fixed and the $(CO)_2$ unit bending out-of-plane. One of these two $A_2$ intermolecular modes (we prefer the former) must be responsible for the 2365.23 cm$^{-1}$ band, and thus has a value of 15.758 cm$^{-1}$.

## 5. Conclusions

In this paper, we have assigned and analyzed rotationally-resolved infrared spectra for a number of weakly-bound trimers and tetramers containing $CO_2$, CO, and $N_2$, and also presented *ab initio* calculations of their structures. There are two families of trimers. The first family, $(CO_2)_2$-CO and $(CO_2)_2$-$N_2$, resembles a $CO_2$ dimer (near planar slipped parallel structure) with the CO or $N_2$ aligned along the dimer symmetry axis. The second trimer family, $CO_2$-$(CO)_2$, $CO_2$-$(N_2)_2$, and $CO_2$-CO-$N_2$, has the CO and/or $N_2$ molecules located in equivalent positions in the equatorial plane of the $CO_2$, pointing approximately at the C atom of the $CO_2$. For the tetramer family, $CO_2$-$(CO)_3$, etc., we take the preceding trimers and add a third CO or $N_2$ to the 'side' of the first two (i.e. not in the $CO_2$ equatorial plane). Interestingly, it is $N_2$ and not CO which prefers to be in this side position.

Calculations indicate that each cluster has a number of distinct structural isomers. In all cases but one, the most stable calculated isomer agrees well with the observed spectrum. For the one exception, which is $CO_2$-$(N_2)_3$, the observed spectrum agrees well with second isomer, which is only very slightly less stable than the first. In order to obtain more complete experimental structural information for the clusters studied here, it would be desirable to measure additional spectra with various isotopic substitutions, preferably high precision pure rotational microwave spectra. The ground state rotational parameters reported here should greatly simplify the search for, and assignment of, such microwave spectra.

**Acknowledgements**

The financial support of the Natural Sciences and Engineering Research Council of Canada is gratefully acknowledged. A.P.C. gratefully acknowledges the financial support by University Ca' Foscari Venezia (ADiR funds) and the computing facilities of SCSCF ("Sistema per il Calcolo Scientifico di Ca' Foscari," a multiprocessor cluster system owned by Universita' Ca' Foscari Venezia).